\documentclass[usenatbib]{mnras}

\usepackage[T1]{fontenc}

\usepackage{times}
\usepackage{url}
\usepackage{hyperref}
\usepackage{graphicx}
\usepackage{graphics} 
\usepackage{epsfig}   
\usepackage{microtype}
\usepackage{threeparttable} 
\usepackage{pdflscape}	

\usepackage[]{multibib}
\newcites{App}{SAMPLE REFERENCES}

\def\apj{{ApJ}}                 
                
\def\apjs{ {ApJS}}

\def\aap{ {A\&A}}

\def\mnras{ {MNRAS}}

\def\nat{ {Nature}}

\newcommand{\rxte}{\textit{RXTE}}

\newcommand{\maxi}{\textit{MAXI}}

\newcommand{\be}{\begin{equation}}
\newcommand{\ee}{\end{equation}}

\newcommand{\msun}{{$M_{\odot}$}}

\newcommand{\gtsima}{$\; \buildrel > \over \sim \;$}
\newcommand{\ltsima}{$\; \buildrel < \over \sim \;$}
\newcommand{\prosima}{$\; \buildrel \propto \over \sim \;$}
\newcommand{\gsim}{\lower.5ex\hbox{\gtsima}}
\newcommand{\lsim}{\lower.5ex\hbox{\ltsima}}
\newcommand{\simgt}{\lower.5ex\hbox{\gtsima}}
\newcommand{\simlt}{\lower.5ex\hbox{\ltsima}}
\newcommand{\simpr}{\lower.5ex\hbox{\prosima}}
\newcommand{\es}{erg~s$^{-1}$}

\newcommand{\etal}{{et al.}}

\newcommand{\lx}{$L_{\rm X}$}
\newcommand{\lr}{$L_{\rm r}$}

\newcommand{\lum}{luminosity}
\newcommand{\lums}{luminosities}
\def\ellr{${\ell}_{\rm r}$}
\def\ellx{${\ell}_{\rm X}$}

\title[NS and BH XRB radio:X-rays]{Hard state neutron star and black hole X-ray binaries in the radio:X-ray luminosity plane}

\author[E. Gallo et al.]{
Elena Gallo,$^{1}$\thanks{E-mail: egallo@umich.edu}
Nathalie Degenaar,$^{2}$
and Jakob van den Eijnden$^{2}$
\\
$^{1}$ Department of Astronomy, University of Michigan, 1085 S University, Ann Arbor, MI 48109, USA\\
$^{2}$ Anton Pannekoek Institute for Astronomy, University of Amsterdam, Science Park 904, 1098 XH Amsterdam, The Netherlands
}

\date{Accepted XXX. Received YYY; in original form ZZZ}

\pubyear{2018}
\begin{document}
\label{firstpage}
\pagerange{\pageref{firstpage}--\pageref{lastpage}}
\maketitle
\begin{abstract}
Motivated by the large body of literature around the phenomenological properties of accreting black hole (BH) and neutron star (NS) X-ray binaries in the radio:X-ray luminosity plane, we carry out a comparative regression analysis on 36 BHs and 41 NSs in hard X-ray states, with data over 7 dex in X-ray luminosity for both. The BHs follow a radio to X-ray (logarithmic) luminosity relation with slope $\beta=0.59\pm0.02$, consistent with the NSs' slope ($\beta=0.44^{+0.05}_{-0.04}$) within 2.5$\sigma$. The best-fitting intercept for the BHs significantly exceeds that for the NSs, cementing BHs as more radio loud, by a factor $\sim$22. This discrepancy can not be fully accounted for by the mass or bolometric correction gap, nor by the NS boundary layer contribution to the X-rays, and is likely to reflect physical differences in the accretion flow efficiency, or the jet powering mechanism. Once importance sampling is implemented to account for the different luminosity distributions, the slopes of the non-pulsating and pulsating NS subsamples are formally inconsistent ($>3\sigma$), unless the transitional millisecond pulsars (whose incoherent radio emission mechanism is not firmly established) are excluded from the analysis. We confirm the lack of a robust partitioning of the BH data set into separate luminosity tracks.
\end{abstract}
\begin{keywords}
X-rays: binaries -- stars: black holes -- stars: neutron -- methods: statistical
\end{keywords}


\section{Introduction}

\begin{figure*}
 \vspace{-0.5cm}
	\includegraphics[width=0.7\textwidth]{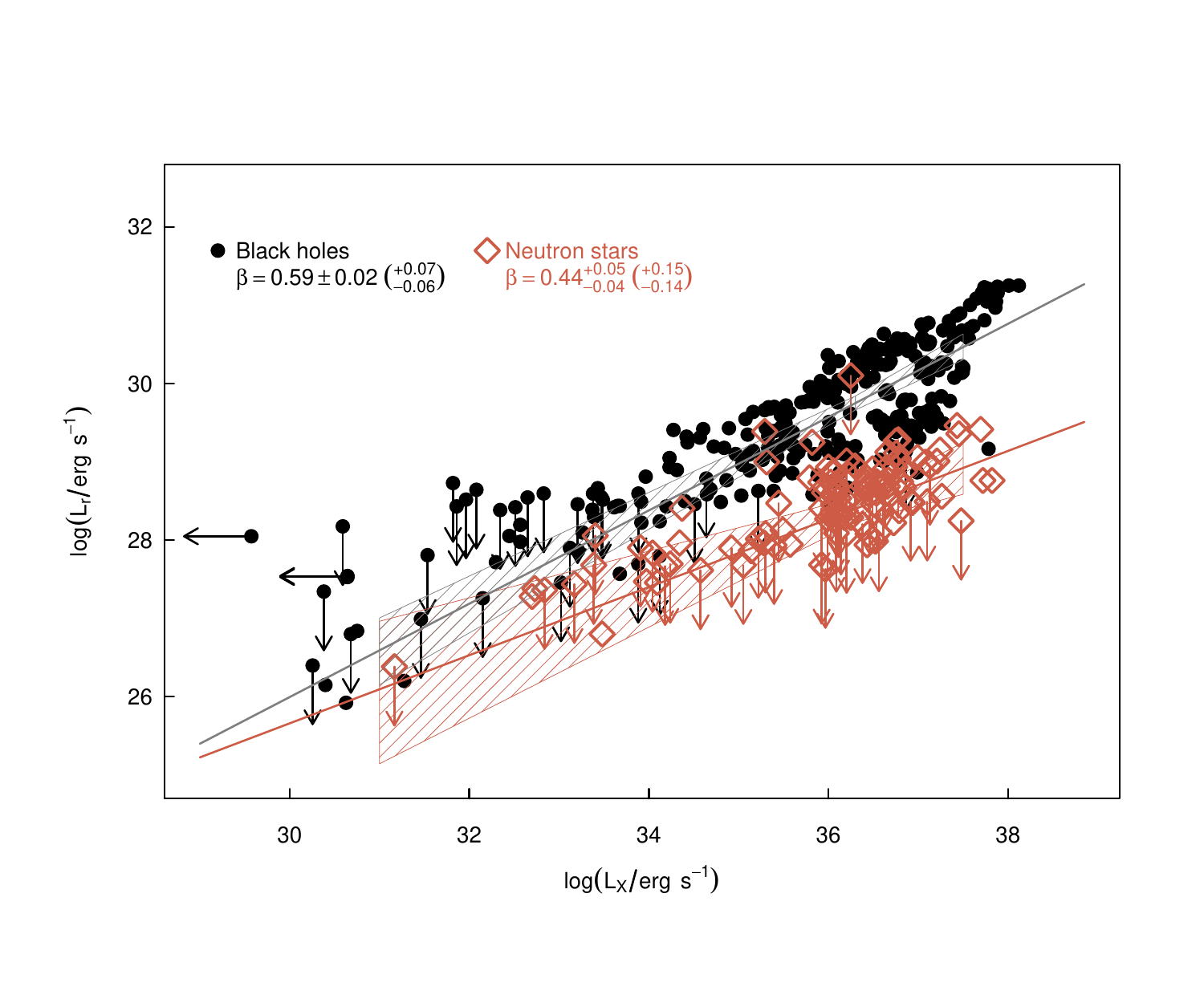}
    \vspace{-0.5cm}
    \caption{Radio (5 GHz) vs. X-ray (1-10 keV) luminosity for a sample of 36 BHs (dark gray circles) and 41 NSs (orange diamonds) in hard states. Solid lines represent the median of the slope ($\beta$) posterior distribution, with 3$\sigma$ errors enclosed by the shaded areas. Quoted errors on $\beta$ are at the 1 (3) $\sigma$ confidence level. }
    \label{fig:fits}
\end{figure*}
In spite of recent major advancements in our phenomenological understanding of X-ray binary (XRB) outflows, and their connection with the accretion flow behaviour, the launching mechanism of relativistic jets is still a matter of debate (see, e.g., \citealt{fender16} for a recent review).  While XRBs jets are typically studied in the radio band, where their synchrotron emission dominates over any other mechanism, they are likely launched and accelerated within a few 100s of gravitational radii of the compact object \citep{gandhi17}, where the X-ray emission originates.  Coordinated radio and X-ray luminosity studies of outbursting XRBs tend to be biased towards the black hole (BH) population. This is likely due to the combination of two main factors. First, the somewhat slower state transition timescales of BH XRBs make it easier to schedule coordinated space and ground observations. Second, and arguably more important, BH XRBs have long been known to be radio brighter than neutron stars (NSs). Seminal work by \cite{fenderkuulkers} showed that transient BH XRBs are more radio loud (in terms of their radio to X-ray luminosity ratio, where both luminosities were measured at peak) than transient NSs. At around the same time, \cite{fenderhendry} noted that the mean radio luminosity of persistent NS Z sources and BHs were broadly consistent with each other, with the persistent Atolls and the transient X-ray pulsars (again at peak) being fainter by a factor of $\simgt$5-10 (the reader is referred to \citealt{munozd14} for recent, detailed reviews on NS X-ray states and nomenclature, and to \citealt{bellonimotta16} for the BHs).

These early studies were concerned with mean and/or peak radio luminosities. Significant advancement came with the onset of multi-wavelength outburst monitoring campaigns (joint with major correlator/receiver upgrades for the Very Large Array and Australia Telescope Compact Array), where a growing number of systems, albeit primarily BHs, have been monitored aggressively in the radio and X-ray band, often over multiple outbursts, and down to very low Eddington ratios. Thanks to these efforts, a coherent phenomenological picture that links distinct X-ray spectral states to different radio properties has emerged for the BHs \citep{fender06,fender09}. Broadly speaking, the hard state is associated with (flat/inverted spectrum) radio emission arising from a partially self-absorbed synchrotron emitting outflow. Compact, hard state radio sources have indeed been resolved into mas-scale collimated jets in two (bright) BHs (GRS1915+205 and Cyg X-1; \citealt{dhawan00,stirling01}), and possibly a NS (Cir X--1; \citealt{millerjones12_cir}). As a system enters into outburst, and its X-ray luminosity starts to increase, so does the radio luminosity of the compact jet. Based on data compiled by \cite{hannikainen98} and \cite{corbel00}, \cite{corbel03} first reported on a tight, non linear correlation between the radio and X-ray luminosity of the BH XRB GX339--4 over about eight years covering three major outbursts: \lx$\propto$\lr$^{0.7}$, over about three orders of magnitude in \lx. Shortly thereafter, based on quasi simultaneous radio and X-ray observations of nine more systems, \cite{gfp03} argued for a universal radio:X-ray luminosity correlation in the hard state of BH XRBs. 

On the NS front, \cite{migliari06} conducted the first systematic investigation, including data from Z sources, accreting millisecond X-ray pulsars (AMXPs), and Atoll systems, both in hard and soft states. For the two (Atoll) systems with hard state data (namely, Aql X-1 and 4U 1728--34) they found \lx$\propto$\lr$^{1.4\pm0.2}$, yielding a significantly steeper correlation than found for the BHs, albeit over a significantly narrower dynamic range. 
Secondly, they confirmed that, whether in absolute or Eddington-scaled units, the "the NSs remain stubbornly less radio loud than the BHs for a given X-ray luminosity", by a factor $\sim$30.   
\begin{table}
\caption{The dependence of radio luminosity upon X-ray luminosity is parametrized as $({\ell}_{\rm r}-28.57)=\alpha+\beta({\ell}_{\rm X}-36.30)$, where $\ell$ denotes logarithmic values in CGS units. 
The best-fitting intercept ($\alpha$), slope ($\beta$), and intrinsic scatter ($\sigma_0$) are defined as the median of the posterior distribution. Errors are quoted at the 1$\sigma$ c.l.; for $\beta$, 3$\sigma$ errors are also given in brackets. \label{tab:fits}}
\begin{threeparttable}
\begin{tabular}{llll}
\hline
Sample &  $\alpha$	  & $\beta$  & $\sigma_0$ \\
\hline 
BH 			& $+1.18^{+0.03}_{-0.03}$ 	& $0.59^{+0.02}_{-0.02}$ ($^{+0.07}_{-0.06}$)			& $0.46^{+0.02}_{-0.02}$ \\
NS 			& $-0.17^{+0.05}_{-0.05}$ 	& $0.44^{+0.05}_{-0.04}$ ($^{+0.15}_{-0.13}$)			& $0.43^{+0.05}_{-0.04}$\\
NS-tMSP 	& $-0.18^{+0.05}_{-0.05}$ 	& $0.49^{+0.06}_{-0.05}$ ($^{+0.22}_{-0.15}$)			& $0.41^{+0.05}_{-0.04}$\\
Atoll			& $-0.25^{+0.06}_{-0.06}$ 	& $0.71^{+0.11}_{-0.09}$ ($^{+0.36}_{-0.24}$)			& $0.30^{+0.05}_{-0.04}$  \\
AMXP 			& $-0.28^{+0.14}_{-0.15}$ 	& $0.27^{+0.09}_{-0.10}$ ($^{+0.29}_{-0.33}$)			& $0.59^{+0.12}_{-0.09}$ \\
AMXP-tMSP 	& $-0.33^{+0.14}_{-0.16}$ 	& $0.19^{+0.13}_{-0.14}$ ($^{+0.45}_{-0.50}$)			& $0.56^{+0.14}_{-0.10}$\\
w1 Atoll 		& $-0.23^{+0.09}_{-0.11}$ 	& $1.16^{+0.28}_{-0.24}$ ($^{+0.91}_{-0.59}$)			& $0.29^{+0.12}_{-0.10}$ \\
w2 Atoll 		& $-0.18^{+0.07}_{-0.08}$ 	& $1.39^{+0.35}_{-0.30}$ ($^{+1.36}_{-1.02}$) 			& $0.21^{+0.11}_{-0.09}$\\
\hline
\end{tabular}
\end{threeparttable}
\end{table}
\begin{figure*}
	\includegraphics[width=0.7\textwidth]{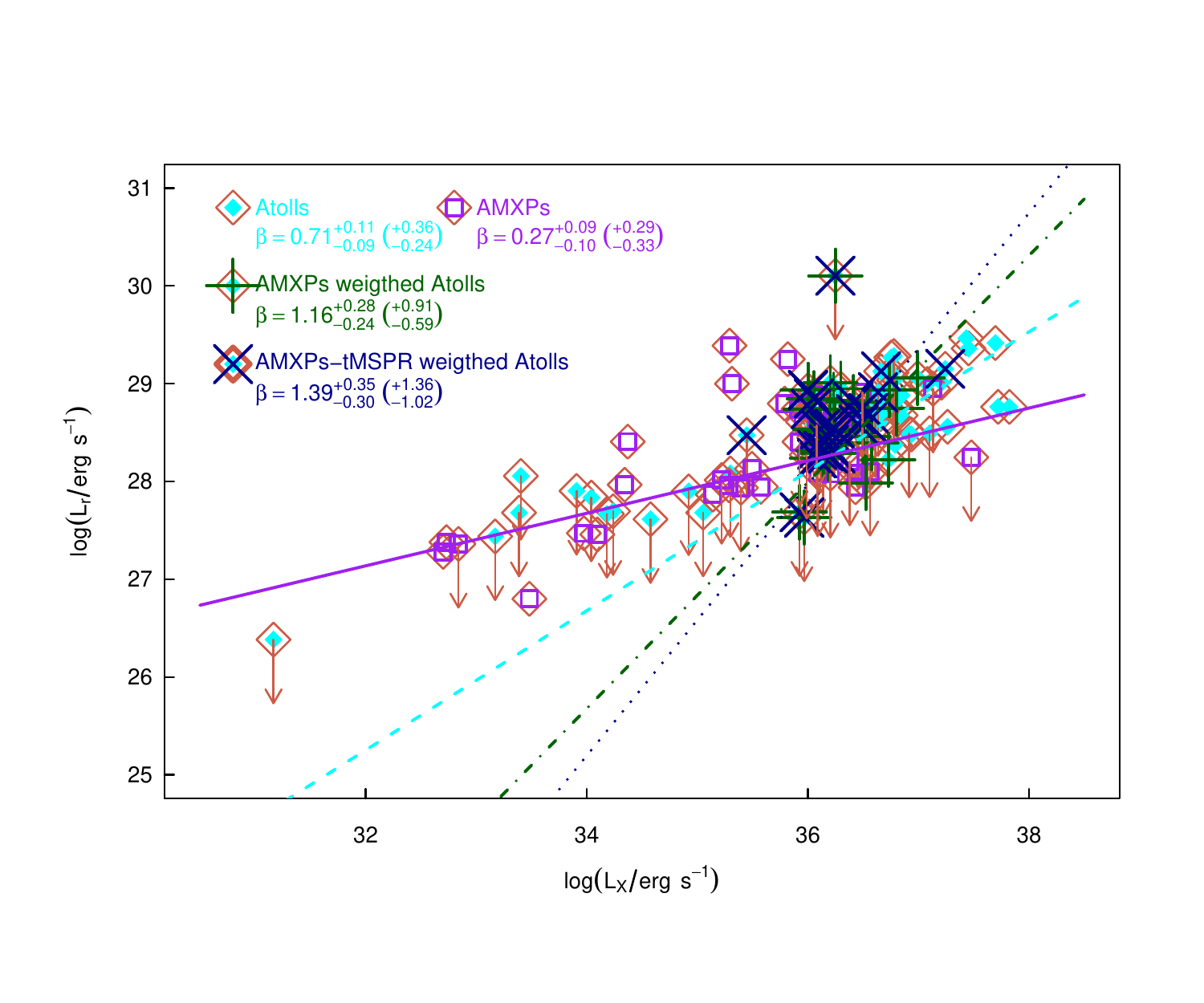}
    \vspace{-0.5cm}
    \caption{\lx\ distribution-weighted NS regression analysis, with weighting functions defined in Fig. \ref{fig:hist}. Quoted errors on $\beta$ are at the 1 (3) $\sigma$ confidence level.}
    \label{fig:weigh}
\end{figure*}
For the reasons discussed above, NS XRB studies have been hampered by the lack of radio detections at X-ray luminosities below $\simlt 10^{36}$ \es. With the exception of the so-called transitional millisecond pulsars \citep[tMSPs;][]{archibald09,papitto13}, for which a similar correlation to hard state BHs has been claimed over more than three orders of magnitude in \lx\ (\citealt{deller15}; see, however, \citealt{bogdanov18}), the limited \lx\ dynamic range makes it challenging to reliably assess the presence of a luminosity correlation for hard state NSs.  In an effort to extend previous investigations to lower luminosities/accretion rates, \cite{tetarenko16gc} and \cite{tudor17} assembled coordinated radio:X-ray observations of a handful additional hard state systems (including new data for the globular cluster Atoll EXO 1745-248, three AMXPs, and the non-pulsating system Cen X-4, both in quiescence as well as in outburst, plus a re-analysis of 4U~1728--34 and Aql X-1 data). Their results point toward a complex pattern of behaviour, with different systems exhibiting different degrees of disc-jet coupling.

In this Letter, we present the largest yet collection of quasi-simultaneous radio:X-ray observations of hard state BH and NS XRBs. A broader study will be presented in a companion work (Van den Eijnden \etal, in prep., including thermal and very high state BHs, Z sources, soft-state Atolls, and slowly pulsating NSs). We start by collating data points from \cite{gallo14}, for BHs, and \cite{migliari06} plus \cite{migliari11}, for the NSs. Additional data points come from further literature searches and or new data collected by our group. For the NSs, we select data for hard state Atolls and AMXPs, including the 3 known tMSPs.  In the case of reported radio fluxes with no accompanying, dedicated X-ray coverage, X-ray fluxes around the time of the radio observation(s) are obtained from the 1-day averaged all-sky monitors on board the \textit{Rossi X-ray Timing Explorer} (\rxte) or \textit{The Monitor of All-sky X-ray Image} (\maxi). X-ray and radio fluxes are converted to unabsorbed 1-10 keV X-ray \lums\ and 5 GHz \lums, respectively. A flat radio spectrum is assumed for the purpose of converting radio flux densities to 5 GHz radio \lum. Overall, our sample is composed of 36 BH and 41 NS systems fainter than \lx$\simlt 10^{38}$ \es; data points, along with a list of adopted distances and references, can be downloaded from: \url{https://jakobvdeijnden.wordpress.com/radioxray/}.

\section{Regression analysis}
We start by assessing whether the distribution of \ellx\ for the NSs (109 data points) is consistent with being drawn from that of the BHs (306 data points, inclusive of 3 X-ray upper limits), where $\ell$ denotes logarithmic \lums\ in units of \es. 
A two-sample Kolmogorov-Smirnov (KS) test probability $p=0.04$ implies that the two distributions are marginally consistent. For each sample, we parameterize the dependence of radio upon X-ray \lum\ as $({\ell}_{\rm r}-28.57)=\alpha+\beta({\ell}_{\rm X}-36.30)$, with intrinsic random scatter included in the fitting. Centering is based on the median luminosities for the combined sample. We run the Bayesian modeling routine described by \cite{kelly07} and implemented in IDL as \textsc{linmix\_err.pro}, which enables us to include censored data in the analysis. 
In line with previous work \citep{gallo12,gallo14}, we assume 0.15 dex uncertainties in both \ellx\ and \ellr, with Gaussian likelihood functions on \ellr\ and \ellx, and uniform probability below upper limits. 
For the purpose of estimating the most likely values of the correlation intercept ($\alpha$), slope ($\beta$), and intrinsic scatter ($\sigma_0$), we calculate the median of 10,000 draws from the posterior distributions; 1$\sigma$ confidence errors are calculated as the 16-84th percentiles of the posterior distributions (Table \ref{tab:fits}). 
Fig. \ref{fig:fits} illustrates the results from the BH vs. NS regression analysis: the BH data points follow a relation with slope $\beta=0.59 \pm 0.02$, intercept $\alpha=+1.18\pm0.03$, and large intrinsic scatter, $\sigma_0=0.46\pm0.02$. For the NSs, $\beta=0.44^{+0.05}_{-0.04}$, $\alpha=-0.17\pm0.05$, and $\sigma_0=0.43^{+0.05}_{-0.04}$. Whereas the two correlation slopes are consistent with each other within 2.5$\sigma$, the best-fitting BH intercept exceeds that for the NS by $1.35$ dex, confirming the BHs as substantially more radio-loud, by a factor $\sim$22. 

\begin{figure}
\vspace{-1cm}
	\includegraphics[width=0.475\textwidth]{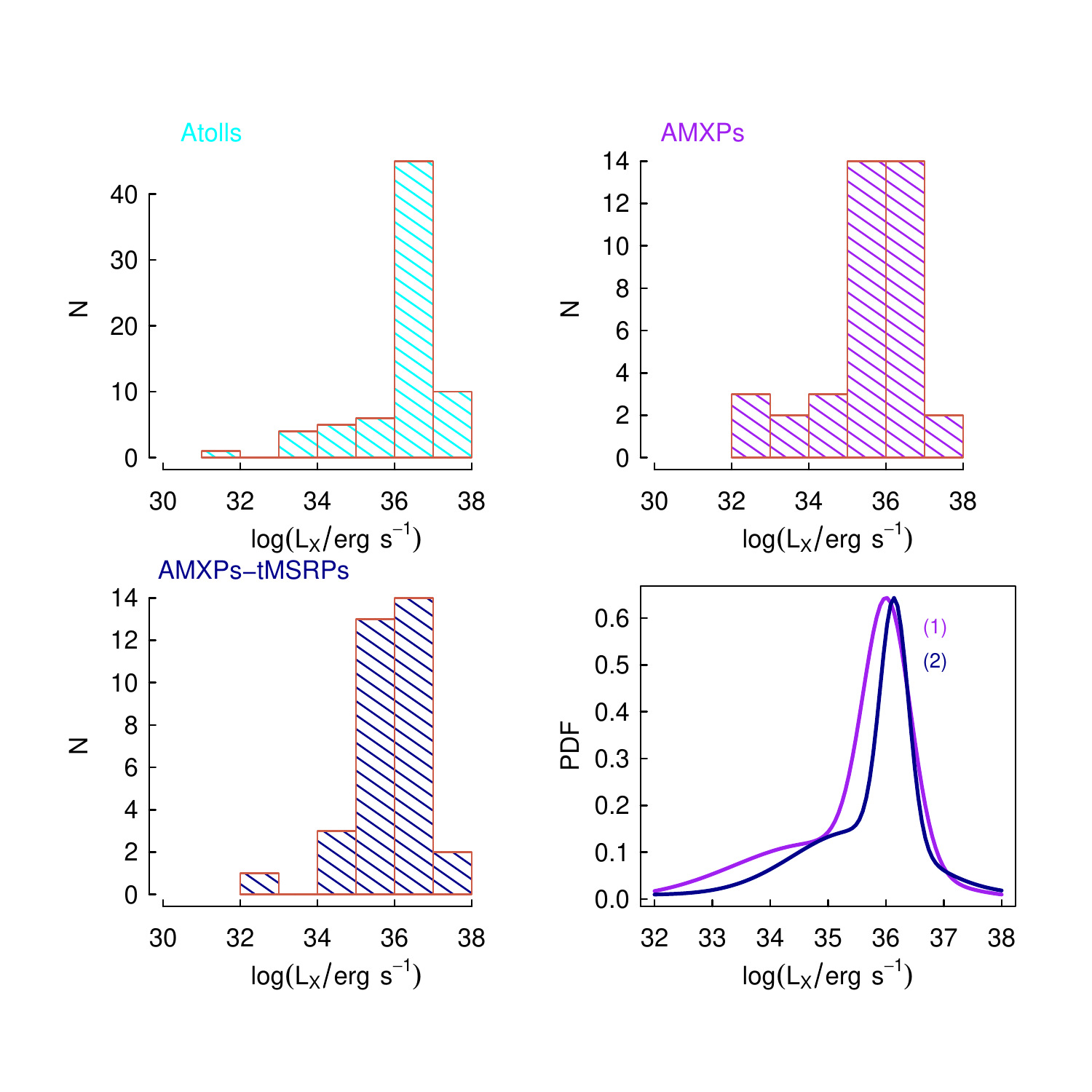}
    \vspace{-0.75cm}
    \caption{X-ray luminosity distributions for the different NS sub-populations: Atolls, AMXPs and tMSPs. The comparative regression analysis is carried out through importance sampling, i.e. using the AMXP (1) and AMXP-tMSP (2) probability density distributions (PDF) as weighing functions.  }
    \label{fig:hist}
\end{figure}

A separate question we wish to address is whether there is any statistically significant difference among different classes of NSs. We start by comparing the sub-sample of X-ray pulsating systems (i.e. the AMXPs, including the 3 tMSPs, for a total of 38 data points) to the (non-pulsating) Atolls (71 data points).
Taken at face value, AMXPs are best described by a shallow relation, with slope $\beta=0.27^{+0.09}_{-0.10}$, whereas $\beta=0.71^{+0.11}_{-0.09}$ for the Atolls (solid purple and dashed cyan lines, respectively, in Fig \ref{fig:weigh}). Although the slopes are consistent within $3\sigma$ (see Table 1), this comparison is intrinsically flawed. A two-sample KS test rules out with high confidence that the two \ellx\ distributions are consistent with each other. 
To control for this, we conduct a weighted comparison by drawing random sub-samples from the Atoll population, weighted to match the AMXP probability density distributions (PDFs) as weighting functions (Fig. \ref{fig:hist}), including and excluding the tMSPs.  
The AMXP-weighted Atoll sub-sample (denoted as w1-Atolls in Table 1) yields $\beta= 1.16^{+0.28}_{-0.24}$ (dashed-dotted green line in Fig. \ref{fig:weigh}), which is formally inconsistent ($>3\sigma$) with the AMXP's slope (the significantly steeper slope of the AMXP-weighted Atolls with respect to the full Atoll sample can be understood noting that the majority of the data points at \ellx$\simgt 36$ are excluded by the importance sampling).  
The same exercise is carried out excluding the 3 tMSPs (5 data points) from the AMXPs population.  Controlling for the different \ellx\ distribution as above (i.e. using PDF 2 in Fig. \ref{fig:hist} as a weighing function), we find that the (AMXP-tMSP)-weighted Atoll distribution (w2 Atolls in Table 1) is best described by $\beta=1.39^{+0.35}_{-0.30}$ (dotted blue line in Fig. \ref{fig:weigh}), which is consistent with the (AMXP-tMSP)'s slope ($0.19^{+0.13}_{-0.14}$) within 2.5$\sigma$. 
 
This begs the question whether the NS vs. BH comparison, too, is affected by the inclusion of the tMSPs. A two-sample KS test shows that, after the exclusion of tMSPs from the NS sample, the NS and BH \ellx\ distributions are still consistent with each other ($p=0.05$), so no importance weighting is required. Re-fitting (NS-tMSP) sample yields a best-fitting slope $\beta= 0.49^{+0.06}_{-0.05}$, which is still consistent with the BHs' within 2$\sigma$.   \\

\begin{figure}
\vspace{-0.5cm}
\hspace{-0.3cm}
	\includegraphics[width=0.5\textwidth]{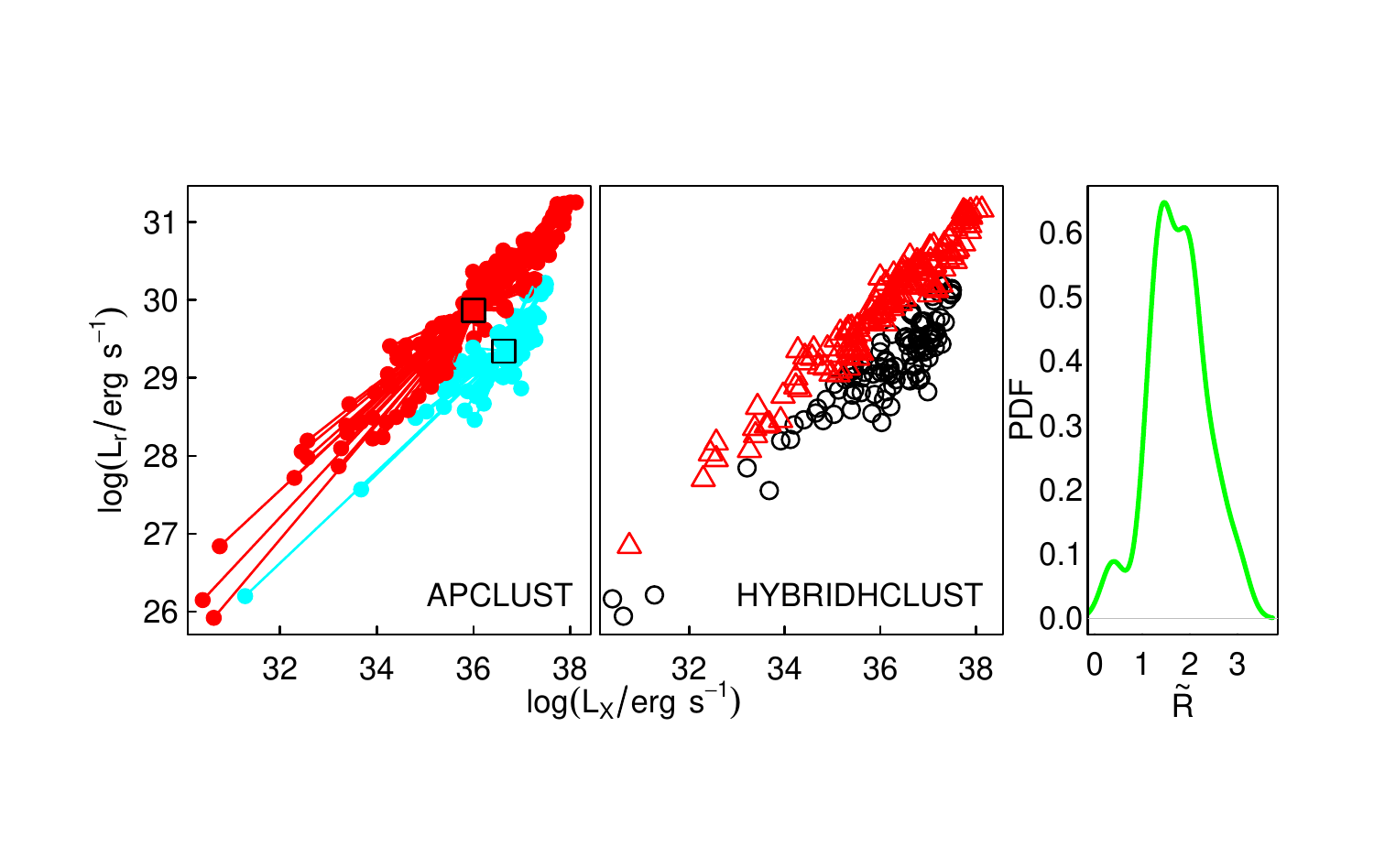}
     \vspace{-0.5cm}
    \caption{Clustering analysis on the BH data points (detections only): both \textsc{APCLUST} (left) and \textsc{HYBRIDHCLUST} (centre) return a two cluster partitioning, albeit with slightly different membership. Right: PDF of the median radio loudness for individual systems, calculated above \ellx=36.2. }\label{fig:apclust}
\end{figure}

Last, we apply the same formalism described in \cite{gallo12,gallo14} to assess whether the BH sample is best described by a single vs. two or more clusters of data points (this approach is intrinsically limited to radio and X-ray detections only). After normalizing the axes to standardized coordinates and rotating them to unit variance, we run the clustering algorithms "affinity
propagation" \citep[\textsc{APCLUST};][]{apclust}, and \textsc{HYBRIDHCLUST} \citep{hclust}. At face value, both prefer a two-cluster description of the data (left and middle panel in Fig. \ref{fig:apclust}). Interestingly, both algorithms identify nearly \textit{parallel tracks}, rather than a main track across the entire \ellx\ dynamic range, plus a steeper track at \ellx$\simgt 35$ (see, e.g. \citealt{coriat11}). The actual cluster membership, however, is not robust. Additionally, after scrambling the data points with a Gaussian distribution of width $\sigma\simgt 0.15$, different cluster centres and group memberships are identified. The lack of a reliable partitioning is confirmed by the fact that the median radio loudness (here defined as \ellr$-$\ellx) for individual systems (calculated above \ellx=36.3, i.e. in the range where two diverging tracks are often claimed), while consistent with a multi-modal distribution (Fig. \ref{fig:apclust}, right panel), does not reflect any of the above clustering analysis results. To summarize, our analysis confirms the lack of a robust partitioning in the BH sample, defined as the ability to reliably and independently define two distinct tracks (this does not, however, constitute proof that the BH data are best described by a single distribution). Owing to this uncertainty, no multi-track regression analysis is warranted for the BH sample. 

\section{Discussion}

As discussed by \cite{tudor17}, individual NS systems exhibit a broad and complex range of behaviour in the radio:X-ray luminosity plane, including anti-correlations (albeit over a limited dynamic range). Additionally, the nature of the incoherent radio emission from tMSPs, i.e. the only NSs with radio detections below \ellx$\simlt 35$, is still a matter of debate \citep{bogdanov18}. Our result that the Atoll and AMXP populations are described by formally consistent scaling relations only if the tMSPs are excluded from the weighted regression analysis goes to show how these kinds of investigations hinge on properly accounting for the different luminosity distributions, and further stresses the importance of unraveling the tMSP radio emission mechanism during the X-ray pulsating mode. 

Regarding the BH sample, we confirm the conclusions by \cite{gallo14} that no {\it robust} partitioning into two or more clusters (or tracks) can be identified. Several authors have investigated the possible origin of the so-called radio-quiet BH track (\citealt{solerifender,dincer14,mh-m14,drappeau15,espinassefender}). While it is entirely possible that the large scatter about the correlation, particularly above $10^{35}$ \es, can indeed arise from an enhanced disc contribution to the X-ray signal, or somewhat steeper radio spectral indices, the formal divide between the so-called standard and radio quiet track is rather blurred, with our latest clustering analysis indicating two nearly parallel tracks. Additionally, in spite of numerous claims to the contrary, the scatter about the NS relation is as large as that of the BHs.  

This study confirms the enhanced radio-loudness of the BH population across a broad dynamic range in X-ray luminosity. The factor of $\simgt$5 difference in the accretor mass can not possibly entirely account for the discrepancy: (naively) correcting for the mass term using the Fundamental Plane of Black Hole Activity relation\footnote{Where we adopt the best-fitting parameters of \cite{merloni03} and use compact object masses from \cite{casares17}, when available, or 2 vs. 10 \msun\ for the NSs and BHs, respectively, otherwise.} yields a difference of 0.81 dex in the best-fitting intercept, implying that the BHs remain more radio loud, by a factor $\sim$6.5, after accounting for the mass term. The gap could be further reduced if the total (as opposed to 1-10 keV) accretion \lum\ were considered: such bolometric correction can be as high as 5--8 for the BHs \citep{zdz04}, vs. a factor 2-3 for the NSs \citep{galloway08}. Adding this to the mass term correction would reduce the radio loudness difference to a factor of 3. Finally, contribution to the quoted 1-10 keV luminosities from the NS boundary layer up to between 30 to 50 per cent of the total X-ray luminosity \citep{burke17} could further reduce the BH-to-NS radio loudness ratio to a minimum of 2.5 (notice that this is obtained by conservatively adopting the largest possible "colour" corrections in all of the above cases). 
This can be equally interpreted as the NSs being significantly brighter in X-rays than the BHs at a given radio luminosity. Advection of energy across the BH event horizon \citep{narayanyi} is unlikely to be responsible for this difference, as one would expect the BHs to become progressively X-ray dimmer with respect to the NSs towards lower X-ray luminosities, which, in turn, would yield inconsistent (at the $>$3$\sigma$ level) slopes. 
Whereas increasing evidence has solidified the discrepancy in radio loudness between BH and NS XRBs, its origin remains unclear. Spin-powering of BH jets remains an appealing explanation, albeit several studies on the matter have failed to reach a compelling conclusion (\citealt{fender10,nmc12,russellspin,steiner13}). Additionally, it has to be noted that the radio luminosity (in and of itself a poor indicator of total jet power) of NSs in soft X-ray states can approach that of the BHs. This will be further investigated in a companion paper (Van den Eijnden et al., in prep.).

\section*{Acknowledgements}
ND and JvdE are supported by a Vidi grant from NWO awarded to ND. We are grateful to James Miller-Jones, Rich Plotkin and Tom Russell for sharing their radio data. We thank the referee for a prompt and insightful report. 


\end{document}